\documentclass[%
 reprint,
superscriptaddress,
 amsmath,amssymb,
 aps,
pra,
onecolumn,
]{revtex4-2}

\usepackage{graphics}

\usepackage{xr-hyper}

\usepackage[utf8]{inputenc}
\usepackage{amsmath}
\usepackage{graphicx}
\usepackage{braket}
\usepackage{slashed}
\usepackage{sidecap}
\usepackage{float}
\usepackage{hyperref}

\usepackage{calligra}
\usepackage{svg}
\usepackage{rotating}

\newcommand{\tinf}{t^{\textrm{inf}}}
\newcommand{\tinfprime}{t^{\textrm{inf}\prime}}

\newcommand{\tsamp}{t^{\textrm{sampling}}}

\newcommand{\thsamp}{\theta^{\textrm{sampling}}}
\newcommand{\thinf}{\theta^{\textrm{inf}}}
\newcommand{\ksamp}{\kappa^{\textrm{sampling}}}
\newcommand{\kinf}{\kappa^{\textrm{inf}}}
\newcommand{\roff}{r^{\textrm{offspring}}}
\newcommand{\poff}{p^{\textrm{offspring}}}
\newcommand{\Poff}{P^{\textrm{offspring}}}
\newcommand{\Psamp}{P^{\textrm{sampling}}}
\newcommand{\Pinf}{P^{\textrm{inf}}}

\newcommand{\tmut}{t^{\textrm{mutation}}}

\newcommand{\cD}{\mathcal{D}}

\newcommand{\cT}{\mathcal{T}}

\newcommand{\Pgen}{P^{\textrm{genetic}}}
\newcommand{\Ploc}{P^{\textrm{location}}}

\begin{document}

\title{Inference of epidemic networks: the effect of different data types} 
\author{Oscar Fajardo-Fontiveros}
 \affiliation{School of Mathematics and Statistics, The University of Sydney, 2006, NSW, Sydney, Australia} 
\affiliation{Centre for Complex Systems, The University of Sydney, 2006, NSW, Sydney, Australia}
  \author{Carl J. E. Suster}
    \affiliation{Centre for Infectious Diseases and Microbiology--Public Health, Westmead Hospital, Westmead, NSW, Australia}
  \affiliation{Sydney Infectious Diseases Institute, Faculty of Medicine and Health, The University of Sydney, Westmead, NSW, Australia}
  \author{Eduardo G. Altmann}
   \affiliation{School of Mathematics and Statistics, The University of Sydney, 2006, NSW, Sydney, Australia} 
\affiliation{Centre for Complex Systems, The University of Sydney, 2006, NSW, Sydney, Australia}
%

%
\begin{abstract}
We investigate how the properties of epidemic networks change depending  on the availability of different types of data on a disease outbreak. 
This is achieved by introducing mathematical and computational methods that 
estimate the probability of transmission trees by combining generative models that jointly determine the number of infected hosts, the probability of infection between them depending on location and genetic information, and their time of infection and sampling. 
We introduce a suitable Markov Chain Monte Carlo method that we show to sample trees according to their probability. Statistics performed over the sampled trees lead to probabilistic estimations of network properties and other quantities of interest, such as the number of unobserved hosts and the depth of the infection tree. 
We confirm the validity of our approach by comparing the numerical results with analytically solvable examples. Finally, we apply our methodology to data from COVID-19 in Australia. We find that network properties that are important for the management of the outbreak depend sensitively on the type of data used in the inference.
\end{abstract}

\maketitle

\section{Introduction}

Some of the most important results in the study of complex systems have been obtained studying the interplay between network connectivity and the system's dynamical properties. In the case of disease spreading, a traditional approach is to consider dynamical models (e.g., compartmental models like the SIR model) on networks with different topology and a major result is the connection between the epidemic threshold and the degree distribution of random networks~\cite{review-complexnetworks}. In this case, the nodes of the network are individuals (who can be infected or not) and links are interactions between them. Generalizations of this approach consider the co-evolution of the disease spreading and of the information individuals have on it, focusing again on the effects of different topologies of the underlying multi-layer networks~\cite{review-coevolution}.
Here we are also interested in the connection between network properties and disease dynamics, but we shift our focus to the role played by data available on the spreading of the disease. This approach is in line with the broader tendency to employ inferential approaches in network science~\cite{peel_statistical_2022,newman16,hric16,Hyland2021MultilayerTypes,Fajardo-Fontiveros2022} . 

The networks we investigate here are transmission trees, with nodes representing infected individuals and directed links representing who infected who. These epidemic networks are not given or taken as an assumption of social-interactions, as in the previous approaches. Instead, they are inferred from the combination of model and data. Thanks to recent technological advances, data on the spreading of diseases is increasingly available, and include both data on infected individuals and genetic information of the virus. 
The primary interest of our paper is on clarifying the effect of different types of (meta)-data on the topological properties of the inferred networks, characterized by different summary statistics. This problem has been investigated also in many other network contexts, including the problems of community detection and link prediction \cite{newman16,hric16,Fajardo-Fontiveros2022}, node-attribute learning~\cite{peel14}, and clustering in networks of documents \cite{Hyland2021MultilayerTypes}.

Our motivation for addressing these problems is that, during an infectious disease outbreak, establishing an accurate estimate of the disease incidence and epidemic dynamics is crucial to effective management. For instance, knowing the number of undetected hosts can inform testing strategies and inferring the transmission tree of the virus can direct targeted interventions to contain the outbreak. Microbial whole genome sequencing (WGS) increasingly plays a role in supporting epidemiological investigation of outbreaks \cite{Sintchenko2015TheTransmission,Wohl2016GenomicOutbreaks}. This indicates the need to understand the extent to which different types of data impact our knowledge of the epidemic dynamics. This is crucial, for instance, when developing surveillance systems that will collect these data, as it is important to consider how the allocation of finite resources might affect the information those systems will provide to responders. This motivates our study on the effect of different types of (meta)-data on the inference of epidemic networks. 

We consider an outbreak scenario of a communicable disease with a host population in which there is no background community transmission. Furthermore, we assume that the substitution rate of the pathogen is high enough that changes in its genome can be informative of transmission, that a laboratory diagnostic assay exists for the disease, and that there is capacity for microbial WGS. The data available for inference therefore consist of the dates of positive laboratory detections (which is related to the date of infection~\cite{Finney2025}), associated microbial genomes where available, and epidemiological data such as the physical location of cases or their membership in suspected epidemiological clusters. Methods for synthesising such data into an inferential framework make use of detailed models of molecular evolution  \cite{Volz2013ViralPhylodynamics,Featherstone2022EpidemiologicalApplications}
and use Monte Carlo exploration of the joint likelihood of the phylogeny and the epidemiological data to sample transmission trees with the greatest support from the data~\cite{Volz2013ViralPhylodynamics,Featherstone2022EpidemiologicalApplications}. 
Inference of transmission trees for outbreaks from microbial WGS has been investigated in a variety of settings, including farm-to-farm transmission of viruses affecting animals \cite{Ypma2012UnravellingData, Ypma2013GeneticInfluenza} and human to human infectious diseases \cite{Wang2020InferencePhase} among others~\cite{Ypma2013RelatingOutbreaks}. The importance of incorporating epidemiological and genomic data into the inference of transmission trees has been demonstrated in different contexts~\cite{Ypma2012UnravellingData,jombart2014,Carson2024,VanderRoest2025}. While these and other state of the art models focus on the accuracy of the inferred (phylogenetic and transmission) networks, our focus here is on the connection between the type of available data and the topology of the network.

The relevance of the problem and scenario described above, and considered in this paper, is exemplified by the response to the recent COVID-19 pandemic. It saw unprecedented use of WGS, enormous publicly-shared viral sequence datasets, and massive laboratory testing of potential cases. This collective effort proved fertile for the development of statistical methods for inferring transmission and estimating key epidemic parameters~\cite{Cori2024InferenceBeyond}. With increasing disease incidence and finite resource availability, even locations that initially achieved comprehensive genomic surveillance~\cite{Brito2022GlobalSurveillance} and contact tracing were forced to screen and sequence selectively. Key questions emerged about the appropriate depth of sampling for WGS and the most effective sampling strategies considering both diagnostic testing and WGS. Methods to address these questions have been elaborated ranging from statistical power calculations \cite{EuropeanCentreforDiseasePreventionandControl.2021GuidanceMonitoring} to more nuanced calculations \cite{Wohl2023SampleBiases} to detailed agent-based models \cite{Han2023SARS-CoV-2Programs} and optimization models \cite{Rasmussen2025}. These approaches have often focused on the specific challenge of minimising the time to detect emerging pathogen lineages or variants, which indeed has been a key objective for many COVID-19 genomic surveillance programs. Tools are lacking to explore the questions in scenarios where transmission inference is possible, such as at the start of an epidemic of a high-consequence disease before its incidence has outpaced the capacity for active case finding and containment~\cite{Suster2022GuidingDetection}.

\begin{figure}[h!]
    \centering
    \includegraphics[width=0.55\linewidth]{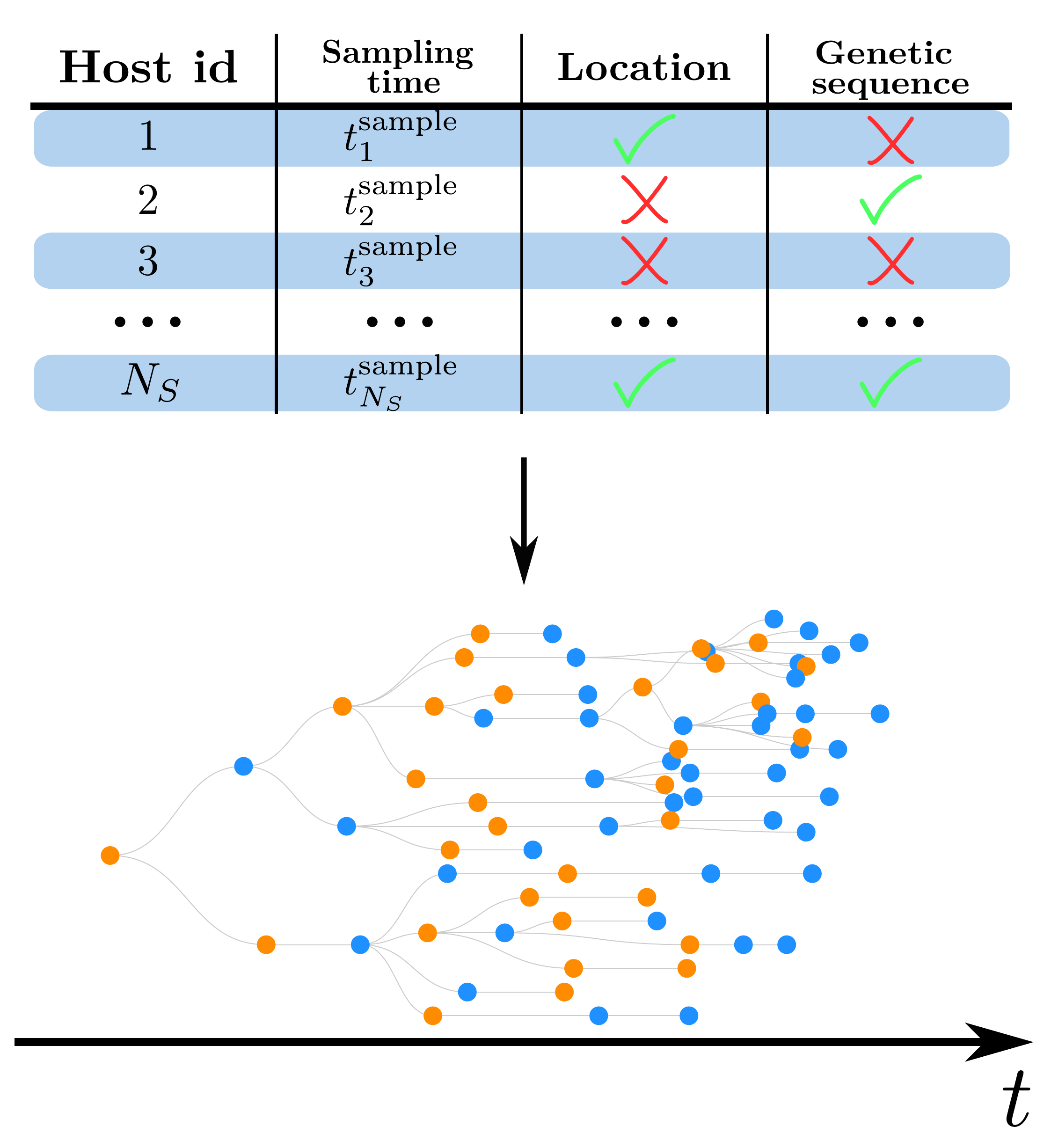}
    \caption{Diverse information about a disease outbreak (top) can be combined to obtain an epidemic network (bottom), which provides a detailed picture of the disease transmission in a population. We consider epidemiological data (sampling time), the pathogen's genome, and the host's location information for $N_S$ sampled hosts. Here we introduce mathematical models and inference techniques that allow us to infer the epidemic network in form of a transmission tree, with blue nodes representing the sampled hosts and orange nodes unsampled hosts.}
    \label{fig:data_to_net}
\end{figure}

In this paper, we introduce an approach for inferring transmission trees $\cT$ using a combination of different types of data~$\cD$ about infected hosts, as illustrated in Figure~\ref{fig:data_to_net}.
This is done by combining different models into a single probabilistic, generative, process of transmission trees.
Each model component is designed to be as simple as possible, making it compatible with a broader range of datasets and suitable for the exploration of the impact of different data types on the inferred trees.
Importantly, the resulting inferred trees can be used to study statistical properties of the topology of these complex networks (e.g., degree distribution, Wiener index) and thus reveal scenarios relevant for outbreak management. For instance, our approach allows for the investigation of hypothetical scenarios for which genomic data are not yet available, in contrast to other approaches that are designed for post-hoc analysis of a specific sequencing dataset.
We propose and test a Metropolis-Hastings Monte Carlo method that we show to sample infection trees according to the probability determined by the combination of data and model.
We validate the model using synthetic data and a set of SARS-CoV-2 genomes from New South Wales (NSW), Australia, collected at the beginning of the pandemic during a period with very high WGS coverage, extensive epidemiological case follow-up, and low incidence.
By comparing the results obtained using different types of data (e.g., location or genomic similarity) separately or in combination, we estimate the effect of additional information on the topology of inferred transmission trees and on the predicted case reporting rate. Our results in the COVID-19 dataset show that inferred trees change substantially with the data, with genetic data leading to transmission trees with a smaller number of unsampled hosts and location data leading to trees with a larger number of unsampled hosts.

\section{Problem Statement}

\begin{figure}[!h]
    \centering
    \includegraphics[width=0.6\linewidth]{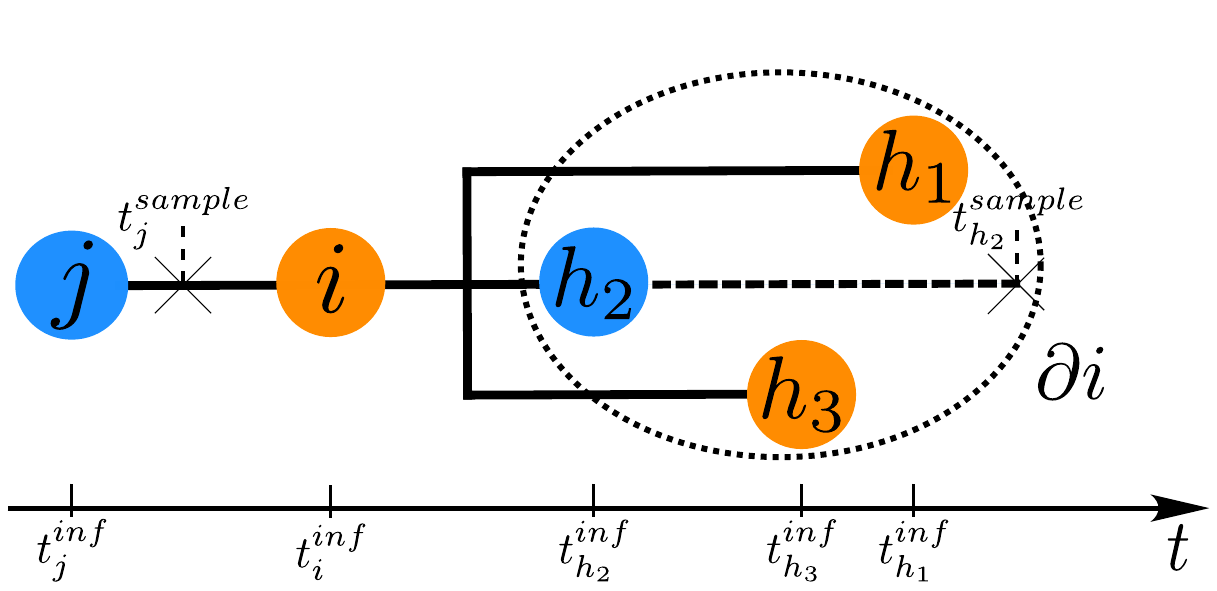}
    \caption{Transmission tree $\cT$ with temporal information. The hosts (nodes) are positioned horizontally according to their infection time $\tinf_i$, inferred by the model. The sampled hosts (in blue) have a known sampling time $\tsamp_i$ (indicated by $\times$) included in the data $\cD$. The number and position of unsampled hosts (in orange) are inferred by the model. The links are directed from left to right and represent transmission events. The nodes directly infected by $i$ are part of the set $\partial i$. }
    \label{fig:simple_case}
\end{figure}

Our aim is to describe the spread of a disease in a population as a transmission tree $\cT$. This tree contains the complete information of how the virus is transmitted through $N$ hosts $i=1, \ldots, N$. For each host $i$ (a node in the tree), $\cT$ contains its infection time $\tinf_i$ and the host $j \neq i$ which transmitted the virus to it (link $j \mapsto i$ in the tree). We denote the set of all hosts directly infected by $i$ as $\partial i$. Figure \ref{fig:simple_case} shows an annotated transmission tree $\cT$.

We will infer $\cT$ from two types of information: characteristics of the virus and the disease it causes (e.g., its mutation rate and how infectious it is), used to determine the parameters of the probabilistic model of disease transmission; and data $\cD$ about hosts in the specific outbreak, used to infer the transmission $\cT$ describing them.  The data $\cD$ contains the following three types of information for a subset of the $N$ hosts (denoted as sampled hosts):
\begin{itemize}
    \item The time $\tsamp_i$ at which they tested positive (sampling time).  
    \item The location of the host (such as testing site or residence). 
    \item The viral genome recovered from the host's sample collected at $t=\tsamp_i$.
\end{itemize}
While we assume that $\tsamp_i$ is available for all $N_S$ sampled hosts, we consider that the location and genetic sequencing may be available only for some of them. Our main interest is to investigate how the properties of the inferred $\cT$ depends on each of the three datasets, i.e., how our knowledge about the transmission of a disease depends on the availability of different information on the outbreak. In Sec.~\ref{sec:model} we introduce a probabilistic model that determines the likelihood of the data $\cD$ being generated by a tree $\cT$, i.e., $P(\cD|\cT)$. In Sec.~\ref{sec.mcmc}, we apply Bayes' formula to compute $P(\cT|\cD)$ and we introduce a Markov Chain Monte Carlo (MCMC) approach to sample trees $\cT$ according to $P(\cT|\cD)$. Finally, in Sec.~\ref{sec:covid} we explore the results obtained applying our approach to data of a COVID-19 outbreak.

\section{Model}
\label{sec:model}

The model we propose focuses on the disease transmission process and ultimately specifies the likelihood $P(\cD|\cT)$ of observing the data $\cD$ for a given tree $\cT$. We consider the observations in each host $i$ conditionally independent of each other (i.e., given $\cT$) so that 
\begin{equation}\label{eq.product}
    P(\cD|\cT) = \prod_{i=1}^N P_i(\cD|\cT).
\end{equation} 
The model for each of the nodes $i$ considers that $P_i(\cD|\cT)$ results from processes modelled through the following five models. Our general approach, and the first three models below, follow Ref.~\cite{Didelot2017GenomicOutbreaks}.

\subsection{Sampling model}

The sampling model specifies the probability that a host $i$ is sampled (tested) and, if it has been sampled, the probability that this occurs at the time $\tsamp_i \ge \tinf_i$. The goal is to capture the varying sampling probability over time such that it is higher at symptom onset and vanishes for $\tsamp_i \gg \tinf_i$ as the host clears the infection. This varying probability depends on $\tsamp_i - \tinf_i$ and is modelled by a gamma distribution $\gamma(t)$~\cite{kroese2013handbook} as 

\begin{equation}
   \Psamp_i 
   =  (1-\pi)^{(1-\sigma_i)}+\pi^{\sigma_i}\gamma(\tsamp_i-\tinf_i;\ksamp,\thsamp),
\label{eq:samp_model}    
\end{equation}
where $\pi$ is the probability that a host is sampled, $\sigma_i=0$ when $i$ is not sampled and $\sigma_i=1$ when $i$ is sampled, and $\ksamp$ and $\thsamp$ are the parameters of $\gamma(t)$. This model does not accommodate persistent infections for which the detection probability can remain high for significantly longer than typical infections~\cite{Machkovech2024PersistentImplications}.

\subsection{Infection model}
The infection model specifies the probability that the transmission to node $i$ at time $\tinf_i$ happened from host $j$. This probability depends on the difference between the infection times of hosts $i$ and $j$, $\tinf_i$ and $\tinf_j$ respectively. This probability will peak at a time $\tinf_i-\tinf_j >0$ when $j$ is most infectious. This process is described by a gamma distribution $\gamma(t)$~\cite{Didelot2014BayesianData}
\begin{equation}
    \Pinf_i =  \gamma(\tinf_i-\tinf_j;k^{inf},\thinf) = \frac{(\tinf_i-\tinf_j)^{\kinf-1}e^{- (\tinf_i-\tinf_j) / \thinf}\left(\thinf\right)^{-\kinf}}{\Gamma(\kinf)},
\label{eq:inf_model}    
\end{equation}
where $\kinf$ is the shape parameter and $\thinf$ is the scale parameter of $\gamma(t)$ and $\Gamma$ is the gamma function~\cite{kroese2013handbook}.

\subsection{Offspring model}

The offspring model specifies the probability of host $i$ infecting $k=k_i$ other individuals (i.e., that node $i$ has an out-degree $k_i$). In order to model the wide variability in $k_i$ observed in many diseases, $\Poff_i$ is described using a negative binomial distribution~\cite{kroese2013handbook} as

\begin{equation}
  \Poff_i 
  = \binom{ k_i + \roff - 1}{k_i} (\poff)^{k_i} (1-\poff)^{\roff},
\end{equation}
where $\roff$ (rate of infection) and $\poff$ (probability of infection) are chosen such that the average $k_i$ is the reproduction number $R$ of the virus.

\subsection{Genetic model}

 Our genetic model is a simplified phylogenetic model in which we assume a constant substitution rate $\mu$. We assume that each infected host has only a single viral population at any time. For each of the (sampled) hosts $i'=1, \ldots, N'$ for which genetic information is available in $\cD$, we consider the closest (sampled) host $j'$ for which genetic information is available and which is not downstream from $i'$ in $\cT$ (i.e., there is no direct path from $i'$ to $j'$)\footnote{Typically, $i'$ will be a descendant of $j'$, i.e., there will be a direct path from $j'$ to $i'$. The exception happens at the top of the tree, when no such node exists and $j'$ is chosen as the closest ``cousin'' (the one which had the shortest time for mutation). This ensures that $P^\textrm{genetic}$ is composed of the same number $N'$ of terms for any tree $\cT$. This is important because the log-likelihood is extensive with the number of data and otherwise we trees $\cT$ with fewer connections between nodes with genetic information would be favoured.}. We define $\Delta \tmut$ as the time that the (single) viral sequence had to mutate between the samplings of $i'$ and $j'$ as  
\begin{equation}
    \Delta \tmut \equiv \tsamp_{i'}-\tinf_{h} + \left| \tsamp_{j'}-\tinf_{h} \right|,
\end{equation}
where $h$ is the first host with no genetic information infected by $j'$ that is a predecessor of $i'$ ($h=i'$ when the link $j'$ is connected directly to $i'$).
The probability that there are $d_{i',j'}$ mutations (measured by single nucleotide polymorphisms, SNPs)  separating the sequences recovered from hosts $i$ and $j$ is then given by
\begin{equation}
    P^{genetic}(j' \to i') = \mu\Delta \tmut e^{-\mu\Delta \tmut d_{i',j'}},
\label{eq:gen_model}    
\end{equation}
where $d_{i,j}$ is the genetic distance. Finally, we take $P_i^\textrm{genetic}=P(j' \to i')$ with $i=i'$ for the nodes $i$ with genetic sequencing and $P_i^\textrm{genetic} = 1$ otherwise.
This simple genetic model constrains the likelihood of transmission trees by a pairwise genomic distance. While less detailed than coalescent models, which use information from the complete sequence, our choice enables the study of different surveillance planning scenarios by varying the substitution rate without needing to specify complete sequences.  This carries the additional benefit of simpler computations and accommodating more diverse datasets (i.e., $d_{i,j}$ can represent other measures of genetic distance besides whole genome SNPs).

\subsection{Location model}

The location model aims to quantify the effect of proximity on the probability of an infection. As a simple case, we consider only whether two hosts $i$ and $j$ are in the same ($\delta_{i,j}=0$) or in different ($\delta_{i,j}=1$) locations\footnote{In the data we study, location is recorded at the level of postal areas.}. As in the genetic model, for each node $i^{\dagger}=1, \ldots, N^{\dagger}$ with location information, we look for the closest host $j^\dagger$ which is not downstream from $i^\dagger$ in $\cT$. If they share the same location ($\delta_{i,j}=0$), we assign a probability of infection $A$. If they do not share the same location ($\delta_{i,j}=1$), we compute the time $\Delta t^\textrm{location}$ between the infections of $i^\dagger$ and $j^\dagger$, where $\Delta t^\textrm{location} \equiv \tinf_{i^\dagger}-\tinf_{j^\dagger}$. If $\Delta t^\textrm{location}$ is small (short time interval between infections), the connection is more unlikely than if $\Delta t^\textrm{location}$ is large (more time between infections, more time for movement).  We assume that the probability of infection between different locations is smaller than $A$ and tends to $A$ as $\Delta t^\textrm{location} \rightarrow \infty$. As a simple functional form that satisfies these constraints, we consider
\begin{equation}\label{eq.location}
    P^{location}(j^\dagger \to i^\dagger) =  \left\{ \begin{matrix}
A \;  \text{if } \; \delta_{i^\dagger,j^\dagger}=0, \\
A(1-e^{-\frac{\Delta t^\textrm{location}}{\tau}} ) \;  \text{if } \; \delta_{i^\dagger,j^\dagger}\neq 0 ,
\end{matrix} \right.
\end{equation}
where $A$ is a constant (fixed by normalization) and $\tau$ is a parameter of the model (time scale for which the probability of infection between different locations differs from the one with the same location). Finally, we consider $P_i^\textrm{location} = P(j^\dagger \to i^\dagger)$ with $i=i^\dagger$ if there is location information on $i$, and $P_i^\textrm{location} =1$ otherwise.


\section{Sampling}
\label{sec.mcmc}

\subsection{Probability of a tree}

We assign to each tree scenario $\cT$ its probability $P(\cT|\cD)$ given the dataset $\cD$. Considering Eq.~(\ref{eq.product}), the independence of the models in Eqs.~(\ref{eq:samp_model})-(\ref{eq.location}), and Bayes theorem, we obtain
\begin{equation}\label{eq.posterior}
    P(\cT|\cD) = P(\cD|\cT) \frac{P(\cT)}{P(\cD)} = \prod_{i=1}^N \Poff_i\Psamp_i\Pinf_i \Pgen_i \Ploc_i \frac{P(\cT)}{P(\cD)},
\end{equation}
where the different $P^\textrm{model}_i$ are given in Eqs.~(\ref{eq:samp_model})-(\ref{eq.location}), $P(\cT)$ is the prior (constant in our analysis), and $P(\cD)$ is the evidence (normalization constant).

\subsection{Markov Chain Monte Carlo (MCMC) approach}
To explore plausible scenarios that explain our dataset $\cD$, we sample trees $\cT$ from the posterior distribution $P(\cT| \cD)$ in Eq.~(\ref{eq.posterior}). This is achieved constructing a Markov Chain Monte Carlo approach that is guaranteed to achieve this goal provided a sufficiently large number of steps $s$ is used. We use the Metropolis-Hastings method that, at each step $s \mapsto s+1$, accepts the change from a network $\cT$ to a new proposed network $\cT'$ with a probability~\cite{kroese2013handbook}
\begin{equation}\label{eq.A}
        A(\cT \to \cT') = \frac{g(\cT' \to \cT)}{g(\cT \to \cT')} \frac{P(\cT'| \cD)}{P(\cT| \cD)},
\end{equation}
where $g(\cT \to \cT')$ is the proposal probability (i.e., the probability of proposing $\cT'$ given the last sampled network $\cT$). For simplicity, here we fix the parameters of the distribution functions appearing in our models ($\thinf, \kinf, \thsamp, \ksamp, \cdots$) using reported epidemiological parameters for COVID-19 (as described in Appendix~\ref{sec.parameters}) \cite{Bar-On2020Sars-cov-2Numbers,Puhach2022SARS-CoV-2Kinetics,McaloonIncubationResearch,Ganyani2020Estimating2020}. In principle, these parameters could be inferred together with $\cT$ by including parameter variations (and their prior probabilities) in the MCMC approach~\cite{Ypma2012UnravellingData}.

\begin{figure}
    \centering
    \includegraphics[width=0.75\linewidth]{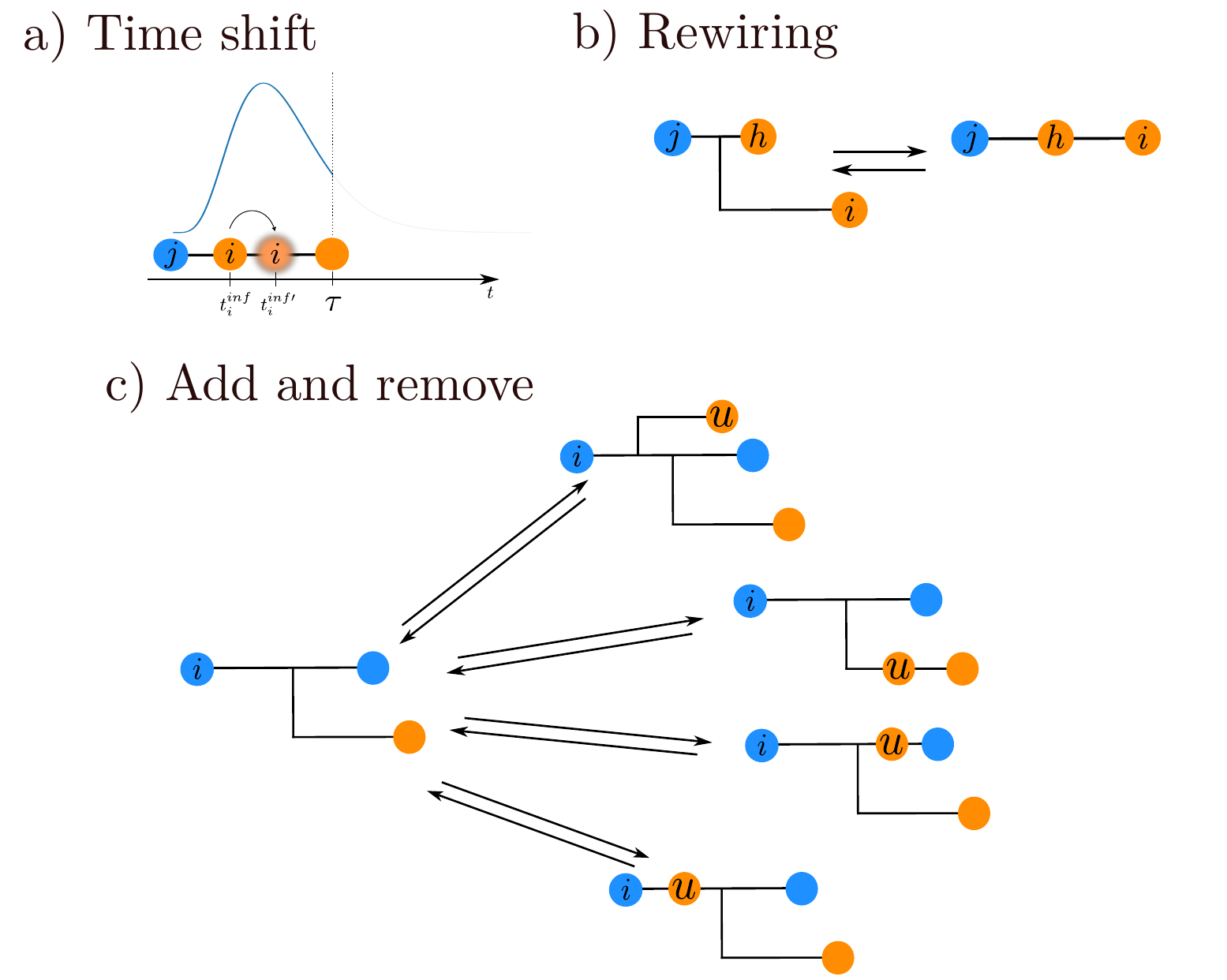}
    \caption{Proposals used in our Monte Carlo sampler. At each Markov step $s$, a new tree $\cT'$ is proposed from the current tree $\cT$, as indicated by arrows ($\leftrightarrow$) in each panel. The proposed tree is obtained performing a local modification to the vicinity of a randomly chosen node $i$ according to one of the following three proposals: (a) time shift of the infecting time $\tinf_i$, Eq.~(\ref{eq.tshift}); (b) rewiring of node $i$ to its grandparent or sibling, Eq.~(\ref{eq.rewiring}); (c) and add (left to right) or remove (right to left) an unsampled host $u$, Eq.~(\ref{eq.add}).}
    \label{fig:proposals}
\end{figure}

The Markov Chain -- constructed applying Eq.~(\ref{eq.A}) at each Markov step $s$ -- samples trees $\cT$ with the desired probability $P(\cT|\cD)$ provided the proposal is such that it is reversible -- i.e., $g(\cT \to \cT^\prime)>0 \Rightarrow g(\cT^\prime \to \cT)>0$ -- and the chain is ergodic -- i.e., there is a non-zero probability of moving from any $\cT$ to any other $\cT^\prime$ as the number of MCMC steps $t$ goes to infinity~\cite{kroese2013handbook}. This requires us to consider variations in the number of unsampled hosts (whose number can grow arbitrarily large), in the links between hosts, and in all infection times. To satisfy these conditions, three proposals $\cT \to \cT'$ with probabilities $g(\cT \to \cT')$  will be applied locally around one host of the tree.  The proposal is thus constructed by first choosing a type of proposal with equal probability ($\frac{1}{3}$), and then choosing one host $i$ uniformly at random from all the hosts for which the chosen proposal can be applied. The three proposals we use are illustrated in Fig.~\ref{fig:proposals}, described in detail below, and implemented in Python in our repository~\cite{Fajardo-Fontiveros_Infectious_tree_network}.

\paragraph{The time shift proposal.} This proposal changes the infection time $\tinf_i$ of a host $i$. The new infection time $\tinfprime_i$ is chosen given by the infection model in Eq.~\ref{eq:inf_model}, truncated by the earliest infection time from $|\partial i|$ and $\tsamp_i$. This is done to increase the performance of the acceptance ratio of the MCMC. The ratio of proposal probabilities of proposing a new infection time $\tinfprime_i$ for host $i$ and returning it to $\tinf_i$ is:

\begin{equation}\label{eq.tshift}
    \frac{g(\cT' \to \cT)}{g(\cT \to \cT')}= \left( \frac{\tinf_{j}  -  \tinf_i}{\tinf_{j}  -  \tinfprime_i } \right)^{\kinf-1} \exp \left(-\frac{\tinfprime_i  - \tinf_i}{\thinf} \right),
\end{equation}
where $\tau$ is the minimum time between $\tsamp_j$ and ${\tinf_k |k \in \partial i}$ and $j$ is the parent of $i$.

\paragraph{Rewiring proposal} We have to differentiate between two types of rewiring to ensure ergodicity. In the first {\it offspring} scenario, host $j$ infects both $i$ and $h$ (see Fig. \ref{fig:proposals}b, left). In the second {\it chain transmission} scenario, $j$ infects $h$ and $h$ infects $i$ (Fig. \ref{fig:proposals}b, right). The proposal has then three steps:

\begin{itemize}
    \item Choose with equal probability ($1/2$) the change scenario to be proposed: from chain to offspring or vice-versa.
    \item Choose with equal probability a host $i$ that can be rewired according to the selected change scenario. We denote by $N_c$ ($N_o$) the number of different hosts for which the  chain to offspring (offspring to chain) scenario can be applied. 
    \item Perform the selected rewiring scenario to the selected node.
\end{itemize}
Taking these steps into account, the ratio of proposal probabilities (``from chain to offspring'' divided by ``from offspring to chain'') is 

\begin{equation}\label{eq.rewiring}
    \frac{g(\cT' \to \cT)}{g(\cT \to \cT')}= \frac{\frac{1}{N_o^\prime} \frac{1}{k_j-1}}{\frac{1}{N_c}},
\end{equation}
where $N_o^\prime$ is the number of nodes that can be rewired from offspring to chain scenario in $\cT'$ and $k_j$ is the out-degree of host $j$.

\paragraph{Add or remove a host proposal.} 
Here we propose how to add or remove an unsampled host $u$ connected to $i$. 
There are in total $2^{k_i}$ ways to connect $i$ to $u$ (see Fig.~\ref{fig:proposals}c for an example with $k_i=2$). We choose among these options with equal probability, leading to the ratio of proposal probabilities 

\begin{equation}\label{eq.add}
    \frac{g(\cT' \to \cT)}{g(\cT \to \cT')}= \left\{\begin{matrix}
                                             \frac{\frac{1}{N_U^\prime}}{\frac{1}{N} }  \frac{1}{\gamma (\tinf_i-\tinf_u;\kinf,\thinf)}\;\; if \;\;   k_i=0\\[16pt] 
                                             \frac{\frac{1}{N_U^\prime}}{\frac{1}{N} \frac{1}{2}}  \frac{1}{\gamma (\tinf_i-\tinf_u;\kinf,\thinf)}\;\;  if \;\;   k_i>0 \text{  and  }  k_u=0\\[16pt] 
                                             \frac{\frac{1}{N_U^\prime}}{\frac{1}{N} \frac{1}{2} \frac{1}{k_i} \frac{1}{\binom{k_i}{k_u}}} \frac{1}{\gamma (\tinf_i-\tinf_u;\kinf,\thinf)}\;\;   if \;\;   k_i>0 \text{  and  }  k_u>0\\
 \\
\end{matrix}\right. ,
\end{equation}
where $N_U^\prime$ is the number of unsampled hosts in the new network with the new unsampled host. The factor $1/2$ is included to account for the probability that $u$ infects (or not) an infected host that was infected by $i$ in $\cT$.

\subsection{Test of MCMC in simple synthetic data}

To test our MCMC sampling, we design a controlled experiment in which the relative probabilities of different tree can be computed exactly and compared to the MCMC sampling. This is obtained by considering data $\cD$ consisting of two sampled hosts, fixing their infection times (i.e., we do not propose another infection time for both of them), and restricting the number of unsampled hosts to at most two (i.e., $N_U \le 2$). This restriction leads to $13$ different classes of transmission networks, each with a fixed connectivity but different possible infection times for the $N_U$ unsampled hosts. With this construction, we ensure that all possible moves in a real case can be applied in this simple case. We compute the (relative) probability of such networks by integrating numerically over the infection time the probability assigned by our model through Eq.~(\ref{eq.posterior}) with parameters fixed as described in Appendix~\ref{sec.parameters}. Figure~\ref{fig:freq_ratio} shows the numerical results for three different networks (depicted in the right). It shows that the ratio of sampled networks obtained through our MCMC scheme converges to the relative probability of the networks computed directly from our model. This result confirms that ratios of probabilities and frequencies coincide for sufficiently long Markov steps $s$.

\begin{figure}[h!]
    \centering
    \includegraphics[width=0.8\linewidth]{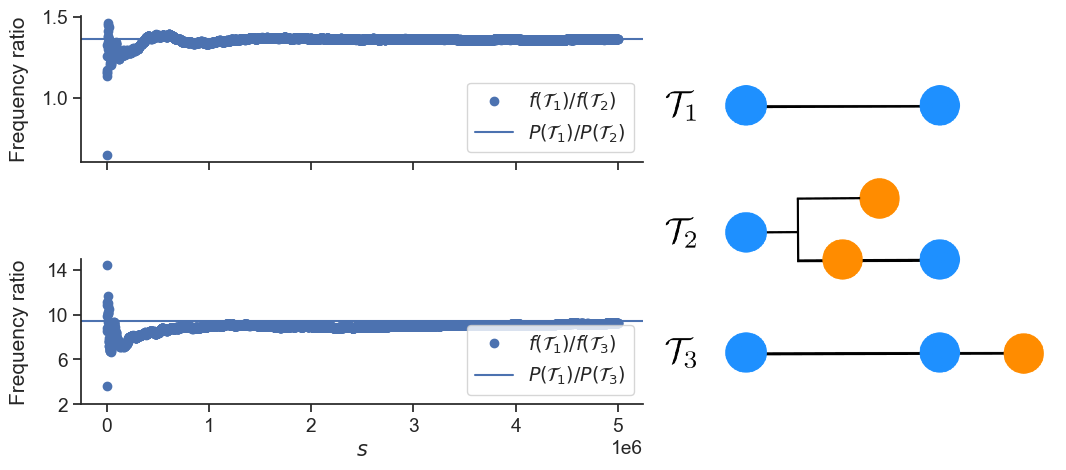}
    \caption{Convergence of the MCMC sampler to the theoretical results in the synthetic data. Each plot shows the ratio of the frequency of two networks $\cT$ (see legend and networks $\cT_1, \cT_2$, and $\cT_3$ on the right) as a function of the number of Markov steps performed. The straight lines are the theoretical result computed by integrating numerically $P(\cT|\cD)$, as described in the text. The data $\cD$ corresponds to the sampling times of $N_S=2$ sampled hosts, generated from our theoretical model. We ran the MCMC described in Sec.~\ref{sec.mcmc} with parameters defined in Appendix~\ref{sec.parameters} for $s=1, 2, \ldots, 5\cdot10^6$ steps. At each step $s$, we compute the frequencies $f(\cT)$ of each structure ($\cT_1, \cT_2,$ and $\cT_3$) in $[0,s]$ and compare their ratio (symbols) to the theoretical expectation (solid line). }
    \label{fig:freq_ratio}
\end{figure}

\section{Application to COVID-19 data}\label{sec:covid}

\subsection{Characterization of the data}

During 2020, the Australian state of New South Wales (NSW) experienced three waves of COVID-19. All three waves were successfully contained through public health interventions, guided by extensive diagnostic screening, detailed case follow-up for active case finding, and viral WGS~\cite{Arnott2021DocumentingAustralia,Capon2021BondiPHRP}. In mid-2021, the Delta SARS-CoV-2 variant of concern was detected circulating in the community in NSW. Disease incidence rapidly outpaced the available resources for comprehensive contact tracing and genomic sequencing \cite{Stobart2022AustraliasCOVID-19}.

The data used in this study are taken from an early study covering the early months of the Delta wave in NSW \cite{Suster2022GuidingDetection}. We select a subset of $N_S=49$ cases with associated SARS-CoV-2 genomes. The data consists of a genetic distance matrix describing the pairwise count of single nucleotide polymorphisms between sequences, the collection dates of the clinical specimens from which the genomes were recovered, and an anonymised pairwise distance matrix of patient addresses (available for $46$ hosts) at the approximate resolution of a postal area. The parameters of the model were fixed based on epidemiological information on COVID-19, as described in Appendix~\ref{sec.parameters}.

\subsection{MCMC sampling}

\begin{figure}[tb]
    \centering
   \includegraphics[width=1\linewidth]{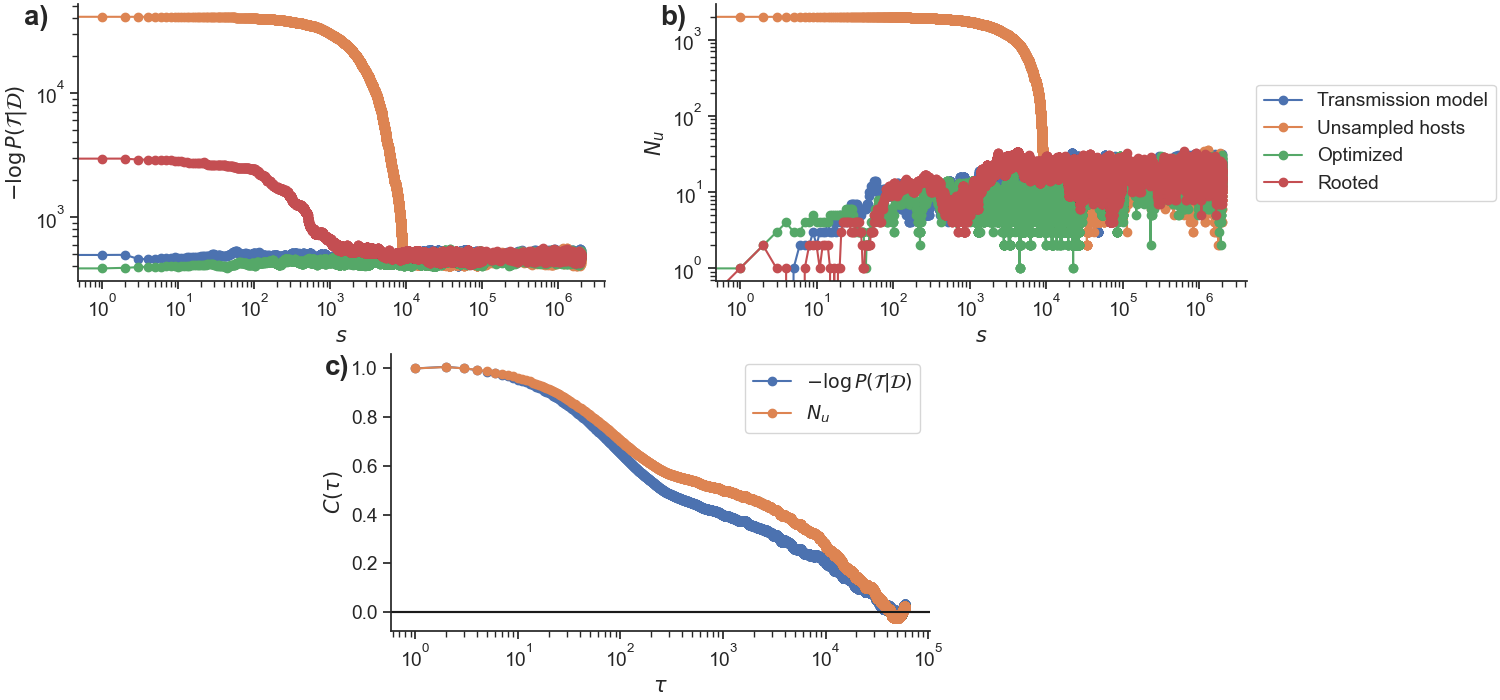}
    \caption{Equilibration of the MCMC in the COVID-19 data. (a) Negative log-posterior  $-\log P(\cT|\cD)$, see Eq.~(\ref{eq.posterior}); and (b) number of unsampled hosts $N_U$ as a function of the number of MCMC steps $s$, for four different initial conditions (see Appendix~\ref{sec.ic} for details). (c) The autocorrelation $C(\tau)$ as a function of the lag time $\tau$ computed for the cases shown in panels (a) and (b), using the samples with $s>10^5$. These results were obtained used as data $\cD$ the sampling time and the genetic distances of $N_S = 50$ sampled hosts.}
    \label{fig:convergence}
\end{figure}

In this section, we quantify the equilibration and mixing properties of our MCMC sampler to ensure that our numerical procedure is sampling trees $\cT$ according to the probability $P(\cT|\cD)$ determined by the data $\cD$ and model.  Starting at an initial condition $\cT_0$ (see Appendix~\ref{sec.ic}), we evolve $\cT$ according to the MCMC obtaining one tree $\cT_s$ for each Markov step $s=0, \ldots, s_{max}$. We then compute different properties $q$ of $\cT_s$, such as their probability $P(\cT|\cD)$ according to our model, the number of unsampled nodes $q=N_U$, and the number of independent trees $q=N_T$ formed by the sampled nodes. For each such property $q$, we look at how $q$ changes with $s$ and we count how many trees $\cT$ in $s \in [s_{min},s_{max}]$ have a specific value of $q$. The theoretical results~\cite{kroese2013handbook} motivating our sampling method in Sec.~\ref{sec.mcmc} guarantee that for any property $q$ of $\cT_s$ and any initial condition $\cT_0$ compatible with $\cD$, the fraction of sampled trees with property $q$ is 
\begin{equation} \label{eq.q}
P(q) = \lim_{s_{max} \rightarrow \infty}\frac{1}{s_{max}-s_{min}}\sum_{s=s_{min}}^{s_{max}} \delta(q(\cT_s)-q) = \int_\cT \delta(q(\cT)-q) P(\cT | \cD) d\cT,
\end{equation}
where $\delta(0)=1$ and $\delta(x)=0$ for $x\neq 0$.

In order to test the theoretical result in Eq.~(\ref{eq.q}), and quantify the equilibration and mixing time of the Markov Chain, we consider as initial conditions $\cT_0$ radically different transmission trees compatible with our data $\cD$ and observe how estimates evolve with the number of Markov steps $s$.
Figure~\ref{fig:convergence} shows the dependence of $q(s) = q(\cT_s)$ for two different observables/values of $q$ and the four initial conditions. We see that after $s\approx 10^4$ Markov steps, the dependence on the initial condition vanishes and $q(s)$ for the four different simulations fluctuate around the same value (in equilibrium).  The auto-correlation function $C(\tau)$ at lag-time $\tau$ (measured in units of Markov steps) for these time series  -- shown in panel (c) -- decays to zero at time $\approx 5 \times 10^4$, suggesting that roughly $10^5$ steps of the Markov chain are needed to obtain a sufficient number of independent samples of $\cT$.

\subsection{Estimated transmission trees}

\begin{figure}[bt!]
    \centering
    \includegraphics[width=0.8\linewidth]{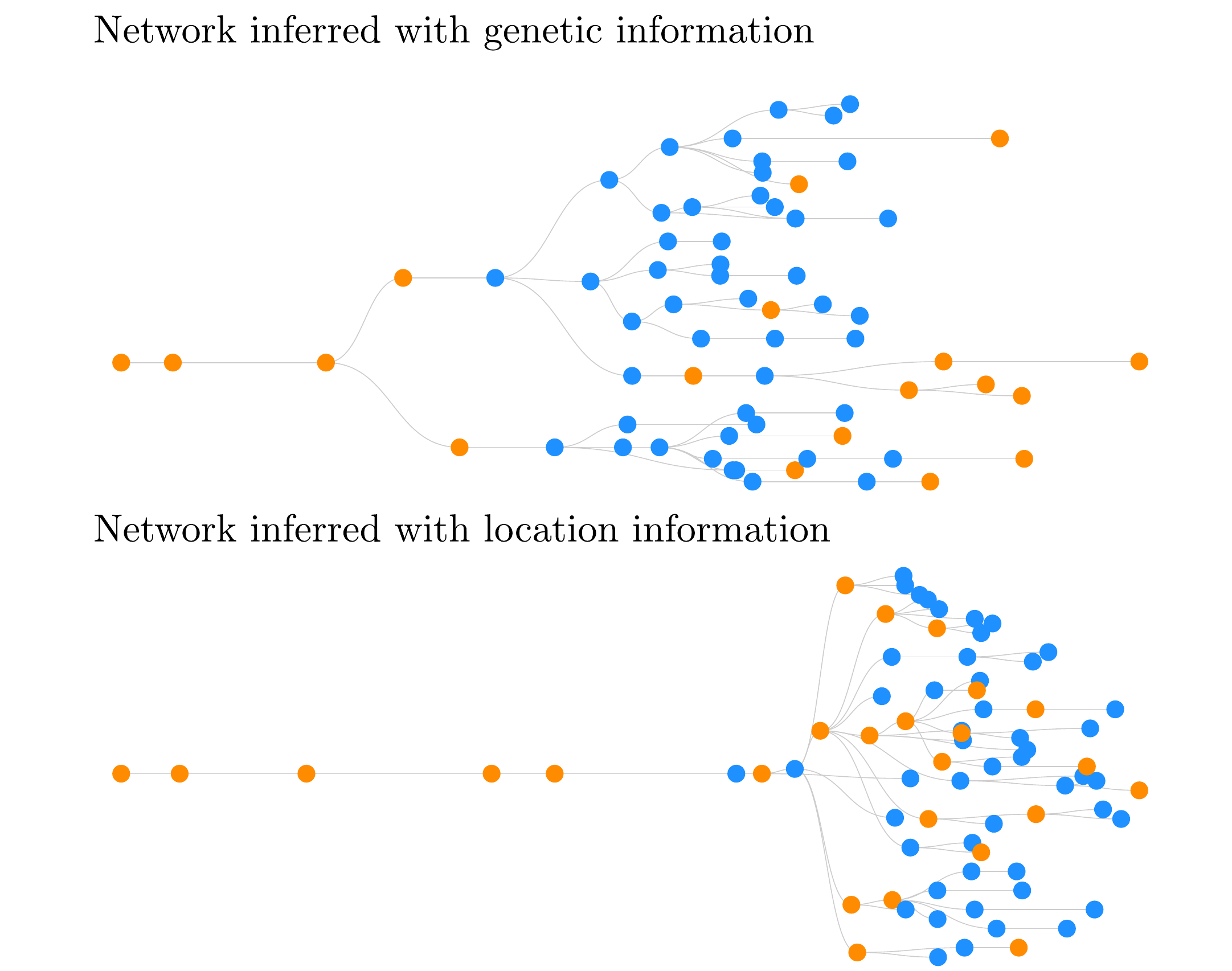}
    \caption{Examples of sampled networks obtained using different datasets $\cD$. Top: transmission tree obtained using as data the sampling times and the genetic information; the sampled tree has $N_U=18$ unsampled hosts and $N_T=2$ subtrees of sampled hosts. Bottom: transmission tree obtained using as data the sampling times and the location information; the sampled tree has $N_U=24$ and $N_T=1$. In both cases, the tree was sampled using the MCMC procedure after equilibration.}
    \label{fig:panel_nets}
\end{figure}

Here we explore the potential of our model and sampling approach by estimating properties of transmission trees for different types of data. Figure~\ref{fig:panel_nets} shows two sampled trees obtained using different datasets. This example suggests that the properties of such trees change substantially depending on the data used in the inference. A key advantage of our methodology is that it estimates for each data $\cD$ not only a single transmission tree, but also different plausible trees. This is done by sampling trees from $P(\cT|\cD)$ and computing statistics over the sampled trees. Based on the convergence properties of our MCMC sampler, we sample $M=1980$ trees in $s\in[s_{min}=10^5,s_{max}= 2\cdot 10^6]$ and explore the probability $P(q)$  in Eq.~(\ref{eq.q}) and the correlation between different observables $q$ of $\cT$.

\begin{figure}[h!]
    \centering
   \includegraphics[width=1\linewidth]{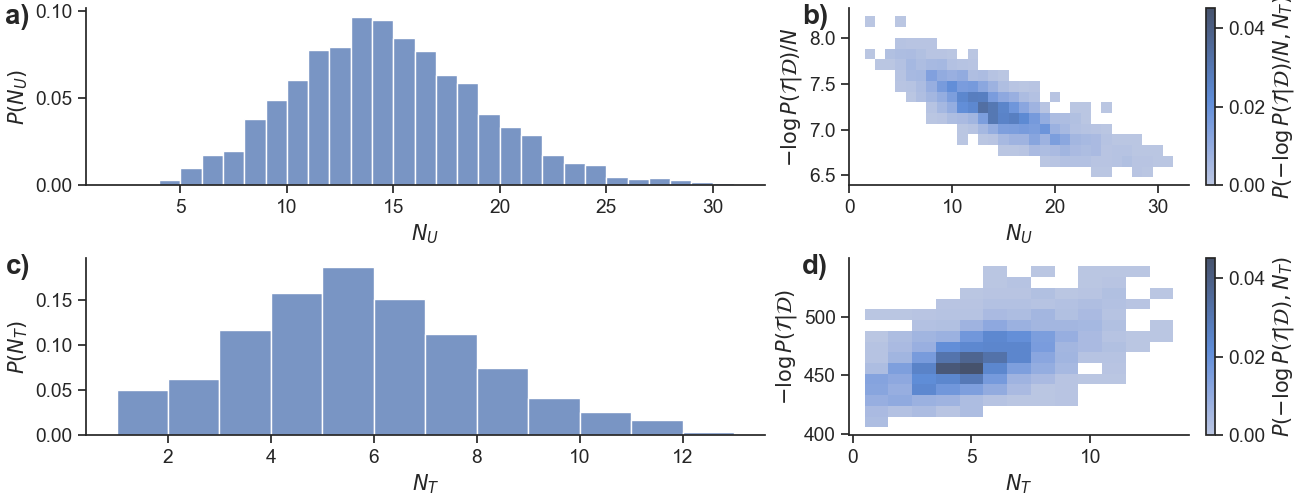}
    \caption{Estimation of quantities of interest computed over plausible transmission trees. (a) Probability of $N_U$ unsampled hosts and (b) joint probability of $N_U$ and the normalized log-posterior (log-posterior divided by the number of hosts). (c) Probability of $N_T$ sampled subtrees  (number of different trees obtained considering only the sampled hosts) and (d) joint probability of $N_T$ and the negative log posterior. $M=1980$ trees $\cT$ were sampled using our MCMC sampling and the COVID-19 data. The probabilities $P(q)$ were computed as described in Eq.~(\ref{eq.q}) for $q=N_U$ and $q=N_T$. See Ref.~\cite{Fajardo-Fontiveros_Infectious_tree_network} for the code used in this figure. }
    \label{fig:histograms}
\end{figure}

Figure~\ref{fig:histograms} illustrates the potential of our methodology. The estimated number of unsampled infected nodes $q=N_U$ varies between $5$ and $25$, with a peak around $N_U=13$. Looking at the correlation between $N_U$ and $P(\cT|\cD)$ in this case, we see that the trees with small $N_U$ have a larger probability $P(\cT|\cD)$. Since there are more possible (and plausible) trees with larger $N_U$, these two effects equilibrate in Eq.~(\ref{eq.q}) leading to a maximum at $N_U \approx 13$.  

\begin{figure}[h!]
    \centering
      \includegraphics[width=0.9\linewidth]{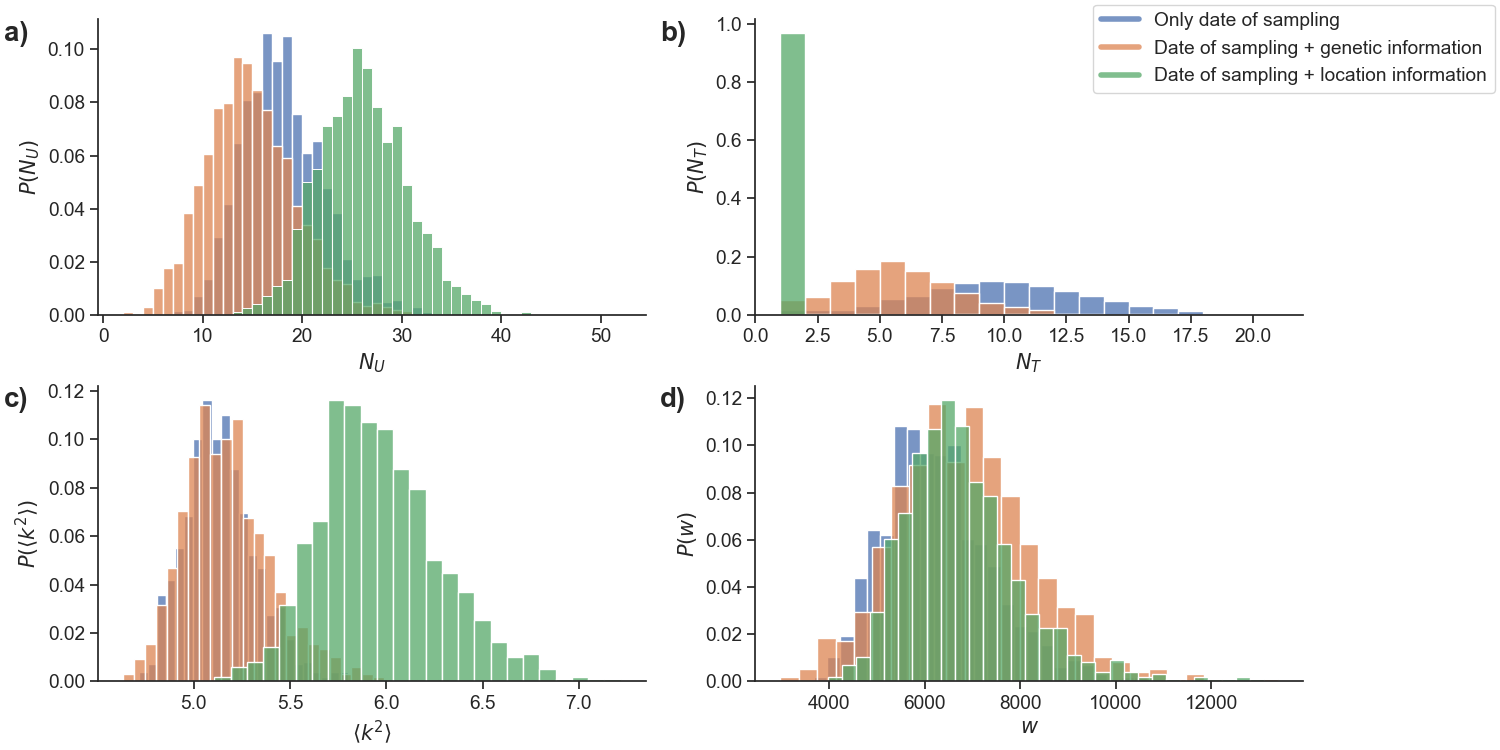}
    \caption{Dependence of transmission trees on the data type $\cD$ used in the inference. Each plot shows the probability $P(q)$ (y-axis) of the transmission tree having observable $q$ ($x$-axis) when different data $\cD$ is used (different colours, see caption). (a) $q=N_U$ number of unsampled hosts; (b) $q=N_T$, number of sampled subtrees (number of different trees obtained considering only the sampled hosts); (c) $q=\left< k^2 \right>$, mean squared out-degree of hosts; (d) $q=W$, the Wiener index defined as the mean pair distances between all nodes~\cite{Goel2015TheDiffusion}. 
    In all cases, $M=1980$ trees $\cT$ were sampled using our MCMC sampling with the COVID-19 data. The probabilities $P(q)$ were computed as described in Eq.~(\ref{eq.q}). See Ref.~\cite{Fajardo-Fontiveros_Infectious_tree_network} for the code used in this figure. }
    \label{fig:histograms_priors}
\end{figure}

Repeating the analysis for different types of data $\cD$, we estimate the extent into which our inference of the transmission tree depends on $\cD$. In this case, we want to compare more measures given the type of data that we have: genetic distance, location information, or nothing. Figure \ref{fig:panel_nets} shows two networks sampled using pairwise genetic distance (top) and location (bottom) information. This would allow us to better understand the effects of extra information on the networks that we sampled. Figure~\ref{fig:histograms_priors} shows how the estimated transmission trees change depending on which types of data (and associated models) are used. For instance, panel (a) shows that the estimated number of unsampled hosts $N_U$ decreases (increases) if genetic distance (location) of the sampled hosts are used to infer the infection trees.

This could change outbreak management measures as the risk of untracked infections will be proportional to $N_U$. More generally,  the results in Figure~\ref{fig:histograms_priors} indicates that the incorporation of meta-data (both location and genetic sequencing) tends to group sampled hosts in fewer subtrees (remarkably, $N_T=1$ for the location case), but location information has an effect on increasing the (variability) of infections per host (out-degree) while genetic information does not. Interestingly, both metadata  have no effect on the estimated Wiener index $W$ of transmission trees, a characterization of trees commonly used in mathematical chemistry, and that has recently been applied in social sciences~\cite{Goel2015TheDiffusion}, defined as the mean pair distances between all nodes (higher $w$, the more viral the virus is).

\section{Conclusions}

We considered the problem of quantifying the dependence of inferred transmission trees $\cT$ of a disease outbreak on different types of datasets $\cD$ of sampled hosts: sampling time, location, and pairwise genetic distance of the virus. Our main methodological contribution is the proposal of a combined model -- Sec.~\ref{sec:model} -- and MCMC computational method -- Sec.~\ref{sec.mcmc} -- that is suitable to address our problem for a variety of datasets and settings. It allows all the data to be used simultaneously to sample transmission trees according to their probability $P(\cT|\cD)$. The  sampling is obtained through a detailed combination of three different MCMC proposals that allow all possible transmission trees to be sampled. We tested the accuracy and convergence of our method in a simple synthetic dataset in which theoretical results could be computed independently. As an illustration of the potential of our general approach in real settings, we considered a simple (representative) dataset of $N_S=49$ sampled hosts during a COVID-19 outbreak in Australia. The results show that when additional data (location or genetic information) is used, the inferred networks show a similar average distance between nodes (as quantified by the Wiener index $W$) but a much narrower degree distribution (as quantified by $\langle k^2\rangle$), confirming a strong connection between available data and the topology of the epidemic network. The implementation of our method in Python is available in our repository, Ref.~\cite{Fajardo-Fontiveros_Infectious_tree_network}.

The main motivation for our investigation is the importance of understanding the impact of different datasets on inferred transmission and surveillance parameters relevant to the management of disease outbreaks. The relevance of this problem is apparent considering settings -- as experienced throughout the early phase of the COVID-19 pandemic in Australia -- in which elimination of outbreaks is targeted and resources can be allocated to additional testing, sequencing, or contact-tracing efforts to support this objective. The results obtained in our simple dataset confirm the potential of our general approach to tackle this problem. In particular, we obtained a quantitative (probabilistic) estimation of the influence of using genetic distance or location information on the inferred trees $\cT$. This is particularly clear on the estimation of the number of unsampled hosts, which varies from $N_U = 17.55$  -- with 80\% confidence interval $CI_{80\%}=[5,14]$ -- when only the sampling time is used to $N_U = 14.28$ -- $CI_{80\%}=[9,20]$ -- (genetic) and $N_U = 25.68$ -- $CI_{80\%}=[20,32]$-- (location) when additional information is included. 
Estimates of $N_U$ can lead to different public health actions due to the implications about the extent of cryptic community transmission. If $N_U$ is very low, it would suggest existing surveillance measures are adequate to contain the outbreak, whereas a high value would suggest the need for enhanced surveillance or a revision in disease control objectives. Our method allows exploration of the impact on the $N_U$ estimate when available data types are changed.
 More generally, sampling transmission trees allows for quantitative estimations of the probability of any epidemic parameter related to the topology of transmission trees and can thus inform data collection strategies aimed at narrowing the uncertainty around estimations scenarios. 

Our approach is based on strong simplifying assumptions that may need to be addressed before considering specific public health applications. Importantly, many of these limitations can be addressed as extension and generalizations of the framework proposed here. For instance, we considered all parameters $\theta$ of our model fixed, while a more realistic setting would be to specify probabilities for different parameter values $P(\theta)$. These probabilities could be used as priors in Eq.~(\ref{eq.posterior})~\cite{Ypma2012UnravellingData}, leading to more accurate estimations of both trees and parameters based on the infection data. Similarly, we considered all positive detections to be real infections while in some settings it might be important to include in our models the possibility of false positive sampled hosts. We also used simplistic genetic and location models, that have the advantage of allowing the application to large classes of (anonymized and aggregated) data but that should be replaced by more accurate models if more detailed data is available. For instance, the genetic model we use assumes a fixed mutation rate $\mu$ and no intra-host viral diversity, while more accurate models of molecular evolution exist and should be used if the full sequence of each case is available~\cite{Volz2013ViralPhylodynamics,Featherstone2022EpidemiologicalApplications,Paredes2024}. Similarly, our newly proposed location model is based simply on whether two sampled hosts were in the same location or not, while more detailed models of human mobility could be used when precise location or additional (contact-tracing) information is available~\cite{Lemey2009}.

While our approach is not strongly pathogen-specific, it is designed to model outbreaks of a virus that spreads directly between hosts with a transmission timescale comparable to the mutation rate. Over longer epidemics, it might be necessary to account for re-infection and immunity, requiring changes in the parameters for the probability of (re-)infection and a more complex representation of transmission networks (e.g., it would contain loops). For a bacterial or highly recombinant pathogen a more nuanced definition of genetic distance would be required as the biological mechanism of evolution is different. Our approach would require substantial modifications to model pathogens with transmission dynamics dissimilar to SARS-CoV-2, for instance where transmission routes other than host-to-host are significant (e.g., for food-, water-, or vector-borne pathogens) or where the incubation period or average interval between substitutions are much larger than the serial interval.

\section*{Acknowledgements}
This study was supported by the Centre for Infectious Diseases and Microbiology -- Public Health via the Prevention Research Support Program of the NSW Ministry of Health. We acknowledge the Microbial Genomics Reference Laboratory, ICPMR-NSW Health Pathology, which generated the NSW SARS-CoV-2 genomes used.

\section*{Data and Code}

The data and code used in this paper are available in Ref.~\cite{Fajardo-Fontiveros_Infectious_tree_network}.

\appendix

\section*{Appendices}

\section{Choice of parameters} \label{sec.parameters}
We fix the parameters of the models presented in Sec.~\ref{sec:model}  based on reported empirical estimates of transmission dynamics and epidemiological information for COVID-19. For each of the five models, we choose the parameters as follows:

\begin{itemize}

\item For the sampling model, we use the proportion of infected hosts who are asymptomatic to set the probability of not being sampled $1-\pi =0.4$\cite{Puhach2022SARS-CoV-2Kinetics}. For the parameters related to the sampling time $t_i^{sampling}$, we use the incubation period of the virus because we consider that hosts are typically tested around the onset of symptoms. We choose the parameters $\ksamp$ and $\thsamp$ of $\gamma(t)$ such that the peak is at the incubation time of the virus (5 days \cite{Bar-On2020Sars-cov-2Numbers,Puhach2022SARS-CoV-2Kinetics}) and that  99\% of symptomatic infected hosts are tested within 14 days because the viral load (and therefore detection probability) is typically low at 14 days after infection~\cite{Bar-On2020Sars-cov-2Numbers,Puhach2022SARS-CoV-2Kinetics}. The values of the obtained parameters are $\ksamp=5.316 \pm0.001$ and $\thsamp=1.158\pm0.001$.

\item For the infection model, we consider that the interval of maximum infectiousness is 4 days \cite{Bar-On2020Sars-cov-2Numbers} and the peak is at 5 days. Proceeding as in the case of the sampling model, we obtain $\kinf = 5.7\pm 1.2$ and $\thinf = 1.06 \pm 0.17$.

    \item For the offspring model, we use the reproduction number $R_0$ reported for early circulating variants of COVID-19 to be $2.7$ \cite{Bar-On2020Sars-cov-2Numbers}. We choose the parameters $\roff$ and $\poff$ such that the mean of the offspring model is $R_0$ and the probability of infecting from $1$ up to $5$ people is $50\%$. This leads to  $\roff=6 \pm 2, \poff=(7.1 \pm  0.9)\cdot 10^{-1}$.

\item For the genetic model, we use an average substitution rate of $\mu = 0.1065 \pm 0.008$ nucleotides per day~\cite{Singh2021OnSARS-CoV-2,Seemann2020TrackingGenomics}. 

\item For the location model, we use $\tau=14$, corresponding to the $14$ days that a typical host is still infectious~\cite{Bar-On2020Sars-cov-2Numbers,Puhach2022SARS-CoV-2Kinetics}.

\end{itemize}

\section{Choice of initial condition}\label{sec.ic}

For all the sampling procedures, we use as a root host an unsampled host that cannot be removed and rewired, as shown in the networks in Figs.~\ref{fig:data_to_net} and~\ref{fig:panel_nets}. The infection time of this root host is chosen $1.5 \Delta t^*$ earlier than the earliest sampled host, where

\begin{equation}
    \Delta t^ * = 4 \left(  \frac{ \left( 1-\pi\right) \Poff(k=1) \left( \thinf \right) ^{-\kinf}}{\Gamma(\kinf)}  \right)^{-\frac{1}{\kinf-1}},
\end{equation}
is the minimum infection time distance between two hosts for which, when you add an unsampled host, the log-likelihood of the system starts to be positive (without taking into account the genetic distance and the location information).

We consider the following choices of initial conditions $\cT(s=0)$ of our Markov Chain:

\begin{itemize}
    \item[i)] Transmission model: the sampled hosts are connected according to the model of Ref.~\ref{sec:model}, ignoring the genetic and location data. Each sampled host is connect to the sampled host with highest probability. There are no unsampled hosts in addition to the root, $N_U=1$.
    \item[ii)] Unsampled hosts: the same as case i), but we randomly add $N_U=2,000$ unsampled hosts (using the add proposal described in Sec.~\ref{sec.mcmc}). 
    \item[iii)] Optimized: $10^6$ modified MCMC steps are applied to the case i). In each step, a proposal is chosen as described in Sec.~\ref{sec.mcmc}, but the step is only accepted if it increases the transmission (offspring, sampling, and infection) component of the posterior in Eq.~(\ref{eq.posterior}). 
    \item[iv)] Rooted: Here we connect the root (unsampled) host to all the sampled hosts, so that $N_U=0$ and $N_T=N_S$.
\end{itemize}

These cases are deliberately chosen to be significant different from each other, allowing for an investigation of their convergence and the equilibration of the Markov Chain, see Sec.~\ref{sec:covid} and Fig.~\ref{fig:convergence}). Later results used initial condition i) because it showed the fastest convergence.


\begin{thebibliography}{10}

\bibitem{review-complexnetworks}
Pastor-Satorras R, Castellano C, Van~Mieghem P, Vespignani A.
\newblock Epidemic processes in complex networks.
\newblock Rev Mod Phys. 2015;
\newblock 87:925--979.

\bibitem{review-coevolution}
Wang W, Liu QH, Liang J, Hu Y, Zhou T.
\newblock Coevolution spreading in complex networks.
\newblock Physics Reports. 2019;
\newblock 820:1--51.

\bibitem{peel_statistical_2022}
Peel L, Peixoto TP, De~Domenico M.
\newblock Statistical inference links data and theory in network science.
\newblock Nature Communications. 2022;
\newblock 13(1):6794.

\bibitem{newman16}
Newman MEJ, Clauset A.
\newblock {Structure and inference in annotated networks}.
\newblock Nat Comm. 2016;
\newblock 7:11863.

\bibitem{hric16}
Hric D, Peixoto TP, Fortunato S.
\newblock {Network Structure, Metadata, and the Prediction of Missing Nodes and
  Annotations}.
\newblock Phys Rev X. 2016;
\newblock 6(3):31038.

\bibitem{Hyland2021MultilayerTypes}
Hyland CC, Tao Y, Azizi L, Gerlach M, Peixoto TP, Altmann EG.
\newblock {Multilayer networks for text analysis with multiple data types}.
\newblock EPJ Data Science. 2021;
\newblock 10(1):1--16.

\bibitem{Fajardo-Fontiveros2022}
Fajardo-Fontiveros O, Guimer{\`{a}} R, Sales-Pardo M.
\newblock {Node Metadata Can Produce Predictability Crossovers in Network
  Inference Problems}.
\newblock Physical Review X. 2022;
\newblock 12(1):011010.

\bibitem{peel14}
Peel L.
\newblock Active discovery of network roles for predicting the classes of
  network nodes.
\newblock Journal of Complex Networks. 2014;
\newblock 3(3):431--449.

\bibitem{Sintchenko2015TheTransmission}
Sintchenko V, Holmes EC.
\newblock {The role of pathogen genomics in assessing disease transmission}.
\newblock BMJ. 2015;
\newblock 350.

\bibitem{Wohl2016GenomicOutbreaks}
Wohl S, Schaffner SF, Sabeti PC.
\newblock {Genomic Analysis of Viral Outbreaks}.
\newblock Annual Review of Virology. 2016;
\newblock 3(Volume 3, 2016):173--195.

\bibitem{Finney2025}
Finney EE, Lee B, Ahmed SF, Sohail MS, Quadeer AA, McKay MR, et~al..
  Back-projection improves inference from sparsely sampled genomic surveillance
  data
  \newblock bioRxiv [Preprint]. 2025;
 \newblock 2025.06.29.662219 

\bibitem{Volz2013ViralPhylodynamics}
Volz EM, Koelle K, Bedford T.
\newblock {Viral Phylodynamics}.
\newblock PLOS Computational Biology. 2013;
\newblock 9(3):e1002947.

\bibitem{Featherstone2022EpidemiologicalApplications}
Featherstone LA, Zhang JM, Vaughan TG, Duchene S.
\newblock {Epidemiological inference from pathogen genomes: A review of
  phylodynamic models and applications}.
\newblock Virus Evolution. 2022;
\newblock 8(1).

\bibitem{Ypma2012UnravellingData}
Ypma RJF, Bataille AMA, Stegeman A, Koch G, Wallinga J, van Ballegooijen WM.
\newblock {Unravelling transmission trees of infectious diseases by combining
  genetic and epidemiological data}.
\newblock Proceedings of the Royal Society B: Biological Sciences. 2012;
\newblock 279(1728):444--450.

\bibitem{Ypma2013GeneticInfluenza}
Ypma RJF, Jonges M, Bataille A, Stegeman A, Koch G, Van~Boven M, et~al.
\newblock {Genetic Data Provide Evidence for Wind-Mediated Transmission of
  Highly Pathogenic Avian Influenza}.
\newblock The Journal of Infectious Diseases. 2013;
\newblock 207(5):730--735.

\bibitem{Wang2020InferencePhase}
Wang L, Didelot X, Yang J, Wong G, Shi Y, Liu W, et~al.
\newblock {Inference of person-to-person transmission of COVID-19 reveals
  hidden super-spreading events during the early outbreak phase}.
\newblock Nature Communications. 2020;
\newblock 11(1):1--6.

\bibitem{Ypma2013RelatingOutbreaks}
Ypma RJF, van Ballegooijen WM, Wallinga J.
\newblock {Relating phylogenetic trees to transmission trees of infectious
  disease outbreaks}.
\newblock Genetics. 2013;
\newblock 195(3):1055--1062.

\bibitem{jombart2014}
Jombart T, Cori A, Didelot X, Cauchemez S, Fraser C, Ferguson N.
\newblock Bayesian Reconstruction of Disease Outbreaks by Combining
  Epidemiologic and Genomic Data.
\newblock PLOS Computational Biology. 2014;
\newblock 10(1):e1003457.

\bibitem{Carson2024}
Carson J, Keeling M, Ribeca P, Didelot X.
\newblock Incorporating epidemiological data into the genomic analysis of
partially sampled infectious disease outbreaks.
\newblock Molecular Biology and Evolution. 2025;
\newblock 42(4):msaf083.
 
\bibitem{VanderRoest2025}
Van~der Roest BR, Klinkenberg D, Fischer EAJ, Bootsma MCJ, Kretzschmar MEE.
  Phylodynamic inference of the contribution of transmission routes in
  infectious disease outbreaks
  \newblock medRxiv [Preprint]. 2025;
\newblock  2025.06.17.25329759.

\bibitem{Cori2024InferenceBeyond}
Cori A, Kucharski A.
\newblock {Inference of epidemic dynamics in the COVID-19 era and beyond}.
\newblock Epidemics. 2024;
\newblock 48:10078.

\bibitem{Brito2022GlobalSurveillance}
Brito AF, Semenova E, Dudas G, Hassler GW, Kalinich CC, Kraemer MUG, et~al.
\newblock {Global disparities in SARS-CoV-2 genomic surveillance}.
\newblock Nature Communications 2022 13:1. 2022;
\newblock 13(1):1--13.

\bibitem{EuropeanCentreforDiseasePreventionandControl.2021GuidanceMonitoring}
{European Centre for Disease Prevention and Control }.
\newblock {Guidance for representative and targeted genomic SARS-CoV-2
  monitoring Key messages Guidance for representative and targeted genomic
  SARS-CoV-2 monitoring}. 2021;.

\bibitem{Wohl2023SampleBiases}
Wohl S, Lee EC, DiPrete BL, Lessler J.
\newblock {Sample size calculations for pathogen variant surveillance in the
  presence of biological and systematic biases}.
\newblock Cell Reports Medicine. 2023;
\newblock 4(5):101022.

\bibitem{Han2023SARS-CoV-2Programs}
Han AX, Toporowski A, Sacks JA, Perkins MD, Briand S, van Kerkhove M, et~al.
\newblock {SARS-CoV-2 diagnostic testing rates determine the sensitivity of
  genomic surveillance programs}.
\newblock Nature Genetics 2023 55:1. 2023;
\newblock 55(1):26--33.

\bibitem{Rasmussen2025}
Rasmussen DA, Bursell MG, Burkhart F. Optimizing genomic sampling for
  demographic and epidemiological inference with Markov decision processes.
  \newblock bioRxiv  [Preprint]. 2025;
  \newblock 2025.06.30.662264.

\bibitem{Suster2022GuidingDetection}
Suster CJE, Arnott A, Blackwell G, Gall M, Draper J, Martinez E, et~al.
\newblock {Guiding the design of SARS-CoV-2 genomic surveillance by estimating
  the resolution of outbreak detection}.
\newblock Frontiers in Public Health. 2022;
\newblock 10:1004201.

\bibitem{Didelot2017GenomicOutbreaks}
Didelot X, Fraser C, Gardy J, Colijn C, Malik H.
\newblock {Genomic infectious disease epidemiology in partially sampled and
  ongoing outbreaks}.
\newblock Molecular Biology and Evolution. 2017;
\newblock 34(4):997--1007.

\bibitem{kroese2013handbook}
Kroese DP, Taimre T, Botev ZI.
\newblock Handbook of Monte Carlo methods.
\newblock John Wiley \& Sons; 2013.

\bibitem{Machkovech2024PersistentImplications}
Machkovech HM, Hahn AM, Garonzik~Wang J, Grubaugh ND, Halfmann PJ, Johnson MC,
  et~al.
\newblock {Persistent SARS-CoV-2 infection: significance and implications}.
\newblock The Lancet Infectious Diseases. 2024;
\newblock 24(7):e453--e462.

\bibitem{Didelot2014BayesianData}
Didelot X, Gardy J, Colijn C.
\newblock {Bayesian inference of infectious disease transmission from
  whole-genome sequence data}.
\newblock Molecular biology and evolution. 2014;
\newblock 31(7):1869--1879.

\bibitem{Bar-On2020Sars-cov-2Numbers}
Bar-On YM, Flamholz A, Phillips R, Milo R.
\newblock {Sars-cov-2 (Covid-19) by the numbers}.
\newblock eLife. 2020;
\newblock 9:e57309.

\bibitem{Puhach2022SARS-CoV-2Kinetics}
Puhach O, Meyer B, Eckerle I.
\newblock {SARS-CoV-2 viral load and shedding kinetics}.
\newblock Nature Reviews Microbiology. 2022;
\newblock 21:147-161.

\bibitem{McaloonIncubationResearch}
Mcaloon C, Collins A, Hunt K, Barber A, Byrne AW, Butler F, et~al.
\newblock {Incubation period of COVID-19: a rapid systematic review and
  meta-analysis of observational research}.
\newblock BMJ Open. 2020; 10:e039652. 

\bibitem{Ganyani2020Estimating2020}
Ganyani T, Kremer C, Chen D, Torneri A, Faes C, Wallinga J, et~al.
\newblock {Estimating the generation interval for coronavirus disease
  (COVID-19) based on symptom onset data, March 2020}.
\newblock Euro Surveill. 2020;
\newblock 17(25):2000257.

\bibitem{Fajardo-Fontiveros_Infectious_tree_network}
Fajardo-Fontiveros O, Suster CJE, Altmann EG. {Data and Python code used in this paper.}
\newblock Available from:
  \url{https://github.com/oscarcapote/transmission_models} (repository) and \url{https://doi.org/10.5281/zenodo.17029983} (permanent link).

\bibitem{Arnott2021DocumentingAustralia}
Arnott A, Draper J, Rockett RJ, Lam C, Sadsad R, Gall M, et~al.
\newblock {Documenting elimination of co-circulating COVID-19 clusters using
  genomics in New South Wales, Australia}.
\newblock BMC Research Notes. 2021;
\newblock 14(1):1--4.

\bibitem{Capon2021BondiPHRP}
Capon A, Sheppeard V, Gonzalez N, Draper J, Zhu A, Browne M, et~al.
\newblock {Bondi and beyond. Lessons from three waves of COVID-19 from 2020 -
  September 2021, Volume 31, Issue 3 | PHRP}.
\newblock Public Health Res Pract. 2021;
\newblock 31(3).

\bibitem{Stobart2022AustraliasCOVID-19}
Stobart A, Duckett S.
\newblock {Australia's Response to COVID-19}.
\newblock Health Economics, Policy and Law. 2022;
\newblock 17(1):95--106.

\bibitem{Goel2015TheDiffusion}
Goel S, Anderson A, Hofman J, Watts DJ.
\newblock {The Structural Virality of Online Diffusion}.
\newblock Management Science. 2015;
\newblock 62(1):180--196.

\bibitem{Paredes2024}
Paredes MI, Ahmed N, Figgins M, Colizza V, Lemey P, McCrone JT, et~al.
\newblock Underdetected dispersal and extensive local transmission drove the
  2022 mpox epidemic.
\newblock Cell. 2024;
\newblock 187(6):1374--1386.e13.

\bibitem{Lemey2009}
Lemey P, Rambaut A, Drummond AJ, Suchard MA.
\newblock Bayesian Phylogeography Finds Its Roots.
\newblock PLoS Computational Biology. 2009;
\newblock 5(9):e1000520.

\bibitem{Singh2021OnSARS-CoV-2}
Singh D, Yi SV.
\newblock {On the origin and evolution of SARS-CoV-2}.
\newblock Experimental {\&} Molecular Medicine. 2021;
\newblock 53(4):537--547.

\bibitem{Seemann2020TrackingGenomics}
Seemann T, Lane CR, Sherry NL, Duchene S, Gon{\c{c}}alves~da Silva A, Caly L,
  et~al.
\newblock {Tracking the COVID-19 pandemic in Australia using genomics}.
\newblock Nature Communications. 2020;
\newblock 11(1):1--9.

\end{thebibliography}
\end{document}